# Examining key features and platforms of IoT

*Rena Bakhshi[1], Mary Hester[2], Jeroen Schot[3], and Lode Kulik[1]*

[1]The Netherlands eScience Center, Science Park 140, 1098 XG Amsterdam, The Netherlands
[2]SURFnet, Moreelsepark 48, 3511 EP Utrecht, The Netherlands
[3]SURFsara, Science Park 140, 1098 XG Amsterdam, The Netherlands

**Abstract**: *To help facilitate expertise in IoT technologies, NLeSC and SURF worked together on a project focusing on IoT applications and platforms. The information included in this case study show the results of NLeSC and SURF's investigation, examining different features offered by cloud and self-maintained IoT platforms with an overall summary of an IoT architecture.*

## 1 Introduction

Internet of Things (IoT) is a paradigm shift, in which all inanimate and animate 'things', are connected and made intelligent while at the same time are embedded and part of the environment. IoT is an integrated technology composed of collaborative sensing, wireless (opportunistic) networking, pervasive computing, in-situ intelligence, sensor data analytics, and active interaction.

Although not an entirely new concept, it has recently gained much popularity especially because of its adoption in many domains, for example health, real-time monitoring and control, and logistics, and new prediction regarding an explosion in number of connected devices in coming years. Unlike their predecessor, *i.e.*, wireless sensor network applications, IoT applications are not application specific, but domain specific and as such bring heterogeneity (in technology, use, requirements, etc), dynamicity, scale, autonomy, and adaptability challenges to a new dimension.

While currently there exist a number of solutions, architectures and platforms supporting co-creation of IoT eco-systems, the diversity and heterogeneity of technological solutions, application segments, requirements, and use cases make it difficult to identify which platform is the best suitable. The challenge is not only to select a platform that solves the interoperability and unification problem of existing IoT technologies and applications, but also the ones yet unforeseen.

This technical note examines different features offered by cloud and self-maintained IoT platforms with an overall summary of an IoT architecture. It is organized as follows: Section 2 describes a generic architecture of IoT platform and its components. In Section 3, we describe and compare most promising open-source IoT platforms. We conclude the note with recommendations in Section 4.

## 2 The architecture of IoT platform

The term *Internet of Things* (IoT) loosely refers to the number of devices (including vehicles and appliances) interconnected with each other and exchanging data via a so-called *IoT platform*. A careful approach to the architectural design can ensure proper integration of a large variety of devices. We will now describe individual components of a generic IoT platform (see an overall architecture in Figure 1), and discuss different options for realizing each component. The interested reader is referred to other surveys [7, 1, 8, 23] and the references as examples of studies on IoT architecture and IoT taxonomy.

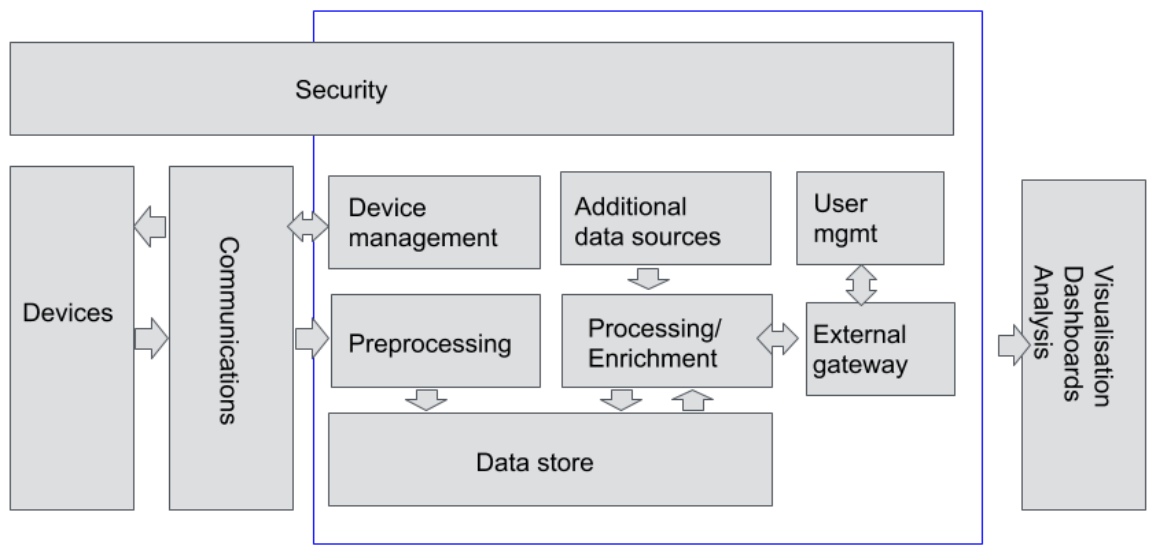

*Figure 1: Schematic depiction of the components of an IoT environment. The blue box depicts the scope of the platform.*

***Device management***

The device management component handles the interaction with the devices. Examples of these interactions are: device registration and activation, monitoring and firmware updates. How to implement this functionality is very dependent on the device types and connectivity. This can make it difficult to provide a generic solution. An advantage is that this component does not interact much with the other components of the platform and is therefore independent.

***Preprocessing***

Preprocessing concerns the transformation of the raw data received from the device before storing it in the *data store*. This transformation can be done for various different reasons. Three common reasons are applying quality control, adding metadata and data restructuring:
- ***Quality control*** Devices might send corrupt data or have sensor malfunctions. During this step we want to detect these malfunctions where possible and either drop the corrupt data or flag the suspicious records so we can decide how to handle this in the *processing/enrichment*.
- ***Adding metadata*** Adding addition meta data to the records is very important for data provenance and can be required for reproducible science. Examples are information about the device (identifier, software version), time received, and version of the preprocessing software.
- **Data restructuring** The data format sent by devices is usually optimized for minimal bandwidth and on-device computation requirements. When storing the data, it is more important to have it structured in a way that allows efficient processing, has good compression and contains schema information/versioning.

There are additional concerns for the preprocessing that depend on the nature of the data transfer between the device and the platform. The first is the grouping of the data: does it received one record at a time, or in batches of data based on time or size? Secondly, is the data flow constant in volume over time, or can there be a sudden peak or large bursts of records?

*Data store*
The data store is responsible for the long-term persistent storage of the data. The two most important aspects of the data store are durability, not losing data after it has been stored, and the ability to handle ever growing data volumes (a form of scalability). Additionally, it is convenient if the data store has good access methods: both efficient querying of subsets of the data and the ability to do parallel reads of large data volumes.

Although we use the singular term of Data Store, this component might include multiple subsystems that each contain either the full dataset or a subset. A common setup uses a scalable file or object store (sometimes called the data lake) for all the raw data as received from the preprocessing component, and one or more databases that contain a subset of the processed/enriched data. These databases are designed and optimized for specific applications.

When choosing the systems used by the data store there are a few different options:
- **File-based** Data is stored in a file-system hierarchy in multiple files. A single file usually contains many records. Data can be accessed via filename, and additional query capabilities are limited. Because of the storage requirements, this is often a distributed file-system where the data is stored on multiple servers and accessible by different clients via a network protocol. Some, but not all, offer a POSIX-like interface to clients to access the data as if it was locally available.
- **Object-based** Similar to file-based storage, but with a flat instead of a hierarchical namespace were just the object label is used to access the data. This has limited capabilities (no in-place editing) with no or limited POSIX-like interface.
- **Database (relational)** In a relational database the data is stored in tables consisting of rows and columns. Relational databases are useful when all rows (also known as records) have the same structure. In practice all relational databases are based on SQL.
- **Database (non-relational)** These are sometimes called NoSQL databases. These databases have different object modes. Examples are document stores, graph databases, key-value stores, column family stores. Often they focus on functioning at a large scale, sacrificing query capabilities or strong consistency to accomplish this.

*Additional data sources*

Often the data stream from sensors or other IoT devices is combined with `static' datasets. These datasets can be part of the research project, or could be from an external party. Examples of these datasets with are the GPS location of all the sensors, or weather information. The platform needs to be able to incorporate these additional datasets and either store a copy or interface with the source data.

*Processing/Enrichment*
Having the raw data available in the data store can be useful, but is often not sufficient. To give meaningful results to the end-user additional processing is needed. This can be simple data processing that only restructures the data, or *data enrichment* where we refine or enhance the data, for example by combining it with additional data sources. There are few different aspects to the processing/enrichment component.

- **Control flow** Control flow can be defined as what triggers the enrichment/processing. There are multiple options that could make sense for separate parts of the processing. This could be event-driven, triggered by new input data; request-driven, triggered by user/API requests; or periodically. The best option depends on the update frequency and if higher latency is acceptable.
- **Storage** The results of the processing can be stored in a database as part of the data store or recomputed on every new user request.
- **Batch/streaming** Depending on the requirements of the application the processing can be done in large batches, or should be using a streaming system.
- **Scalability** It can be the case that a single machine cannot keep up with the processing requirements, as new data keeps coming in and results should be delivered within a short time frame. The processing solution therefore should be scalable in that the work can be distributed over multiple machines. If all records can be processes independent from each other this need not be complicated, but if there are dependencies or aggregations a suitable distributed data processing framework should be used.
- **Validation** There should be a way to check the validity of the data processing, and processing should be annotated to allow the development and improvement in a reproducible fashion.

*External gateway*
We decided that the visualization, dashboards and analysis are out of scope for the core research IoT platform. But these are very important and need a way to interface with the system. This interface (API) is provided by the external gateway. This gateway handles requests from the end-user and returns data based on the request. This data can be be processed before it is returned. The data could be returned as files that the user downloads and processes off-line, or directly handled by a web application. It is important that the API and the structure of the returned data are properly documented.

*User management (external users/researchers)*
Research is never done in isolation, so the external gateway should provide access to end-users from different institutes. We do want to apply some access restrictions, so some form of authentication and authorization is required. Ideally, we do not want to force them to create yet another account but be able to use the credentials from their home institute.

With SURFconext [18], SURF (which is part of the national e-infrastructure for research and education in the Netherlands) offers federative access for academia in the Netherlands.

SURFconext enables single sign-on access to web, cloud, institutional services based on the user's institutional account (and therefore re-using the university identity management user registrations). With millions of authentications per month SURFconext is a very successful solution for any HTTP-based application.

A limitation of SURFconext is the fact that by default it only handles web-based applications. Rich client/non-web applications cannot make easy use of it. If the external gateway is only accessed as a web application, this is not an issue. However, we can imagine some cases where there would be a need for non-web access. A solution for this could be the use of an authorization proxy.

SURF is currently working on a setup of such a proxy, in a project called the Science Collaboration Zone [17], which includes a solution called COmanage [11]. COmanage is a tool that adds a number of useful features, such as on-boarding researchers to one or more virtual collaborative organization (groups) and functionality to register ssh keys to generate one-time passwords and application-specific passwords to enable access to non-web-based resources but all after initial on-boarding based on a verified institutional account.

## 3 Summary of open source candidates

We started this project with the aim to develop a prototype of the IoT platform that works with a wide range of IoT applications as a final deliverable. We were looking for a scalable solution (so able to serve multiple applications/use cases) with minimal changes to the platform, especially with respect to interfaces. To this end, we identified two categories of open-source IoT platforms, *cloud-centric* and *self-maintained*.

Table 1: Comparison of the cloud-centric platforms and their components.

|  | **Amazon Web Services** | **Microsoft Azure** | **Google Cloud Platform** | **IBM Cloud** |
|---|---|---|---|---|
| **Device management** | IoT Platform | IoT Hub | Cloud IoT Core | IoT Platform |
| **Preprocessing (Compute)** | | | | |
| Virtual machines | EC2, Lambda, Kinesis | Virtual Machines | Compute Engine | Bare Metal Servers, Cloud Virtual Servers |
| Containers | Elastic Container Service | Azure Containers Services/Instances | Google Container Engine | Container Service |
| FaaS | Lambda | Functions | Cloud Functions | Cloud Functions |
| Streaming | Kinesis | Stream Analytics | Cloud Dataflow | Streaming Analytics |
| **Data Store** | | | | |
| File/object based | S3, Elastic File System | Blob Storage, Azure Data Lake | Cloud storage | File Storage, Object Storage |
| Database (relational) | RDS, Redshift | SQL Database, SQL Data Warehouse | Cloud SQL, Spanner | IBM Compose (MySQL, PostgreSQL), Db2 Warehouse |
| Database (non-relational) | DynamoDB | CosmosDB, Table Storage | Cloud Bigtable, Datastore | IBM Compose (ScyllaDB) |
| **Additional data sources** | Glue | – | – | Weather Data APIs |
| **Processing/Enrichment** | EMR | HDInsight | Dataproc | Analytics Engine |
| **External gateway** | API Gateway | API Management | Apigee | API Connect |
| **User management** | IAM | Security Center | Cloud IAM | App ID |

Table 2: Comparison of the self-maintained platforms and their components.

|  | Kaa | IoTivity | ThingsBoard | OpenHAB 2 |
|---|---|---|---|---|
| **Device management** | SDK | SDK, Device Management | Tenant Administrator | Paper UI |
| **Preprocessing (Compute)** | | | | |
| Virtual machines | Kaa Sandbox | IoTivity Simulator | – | VM |
| Containers | (Local) Docker container | – | Kubernetes | Ready-made packages |
| Streaming | Apache Spark Streaming | Data transmission | Apache Spark Streaming | REST |
| **Data Store** | | | | |
| File/object based | – | – | – | db4o, RRD4J |
| Database (relational) | PostgreSQL | Resource Data Query Processor | PostgreSQL, HSQLDB | JDBC (PostgreSQL, MySQL), InfluDB |
| Database (non-relational) | MongoDB, Cassandra | Data Management | Cassandra | Amazon DynamoDB, MapDB |
| **Additional data sources** | PubNub Log Appender | Protocol Plugin Manager | Apache Kafka plugin, Sigfox extension | HTTP binding |
| **Processing/Enrichment** | custom modules | Soft Sensor Manager | Rule Engine | Eclipse SmartHome |
| **External gateway** | REST, Apache Flume | IoT REST API Server [10] | REST | "Home" Gateway |
| **User management** | Administration UI | Scene Manager | Tenant Administrator | – |

### *Cloud-centric solutions*

For the cloud-centric IoT platforms, we refer to the recent detailed comparison by Guth et al. [9]. This includes the open-source platforms such as FIWARE[1], OpenMTC, SiteWhere, and Webinos as well as the proprietary solutions such as AWS IoT, IBM's Watson IoT Platform, Microsoft Azure IoT Hub, and Samsung SmartThings. Some of these commercial solutions are surveyed in Table 2.

### *Self-maintained solutions*

For the self-maintained solutions (also known as *on-site* or *on-premise*), we identified four promising open-source community projects: Kaa, IoTivity, ThingsBoard and OpenHAB.

The summary of the survey is presented in Table 2. We discarded one of the criteria from the previous comparison table, FaaS. None of the four platforms offer function platform solutions (FaaS) as a part of the software stack. However, there are a lot of on-premise FaaS that one can embed such as Iron.io (2014), Apache OpenWhisk (2016), Fission (2016), Galactic Fog's Gestalt (2016), OpenLambda (2016), and OpenFaaS (2017).

---

[1] Additional information about FIWARE [2]: It is open-source platform developed out of an EU-funded project (which is now completed). As a cloud-centric solution, it provides a set of standardized APIs to support the creation of smart applications or applications for smart devices (see, for example, Fi-Beer [14] and the Pilot project [16]). Generally speaking, FIWARE is a ``curated framework of open source platform components'' which can be assembled together with other third-party components to facilitate the development of smart applications including FiWare IoT [12].

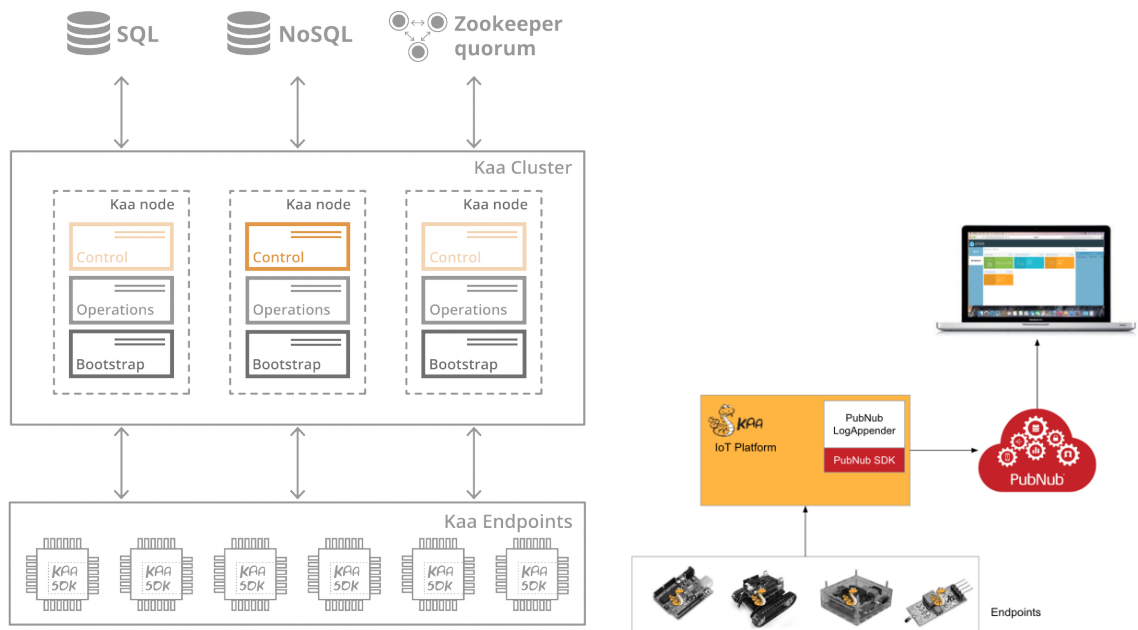

*Figure 2: Kaa IoT platform: Conceptual Architecture (left), and connecting Kaa to Arbela. (Kaa IoT Technologies, https://www.kaaproject.org.)*

**Kaa** [19] is an open-source middleware platform for implementing IoT applications and applications for smart devices. The platform software is easy to install thanks to Kaa Sandbox which is a complete virtual machine image. The sandbox comes with a complete Kaa installation, the sandbox environment, sample applications, three types of databases (PostgreSQL, MongoDB, and Cassandra), Android SDK, and other third-party integration related to enabling different hardware vendors. Fig. 2 (left) depicts a conceptual architecture of Kaa; for more details on the components, we refer to the Kaa documentation (http://kaaproject.github.io/kaa/docs/v0.10.0/Architecture-overview/). It is released under an Apache 2.0 licence via a GitHub repository (https://github.com/kaaproject). Kaa enables collecting data from devices that use PAN-based protocols such as Bluetooth, ZigBee, and Z-Wave. Kaa endpoint software development kits (SDKs) handle client-server communication, authentication, data marshaling, encryption, persistence and other services provided by the Kaa platform. In principle, Kaa can handle both structured and unstructured data, though it can manage devices that share the same set of data schemas (Apache Avro-compatible). Kaa supports a framework of pluggable log appenders (e.g., PubNub Log Appender) in order to load data into a database. The data can be send to stream processing or can be made available to custom data processing modules via REST or Apache Flume. The Kaa Cluster uses Apache Zookeeper for the coordination of servers, Kaa node elections, failure mitigation, and load balancing. To enable real-time monitoring, Arbela [22] can be used as a Kaa IoT Dashboard using the PubNub channel (see Fig. 2 (right)).

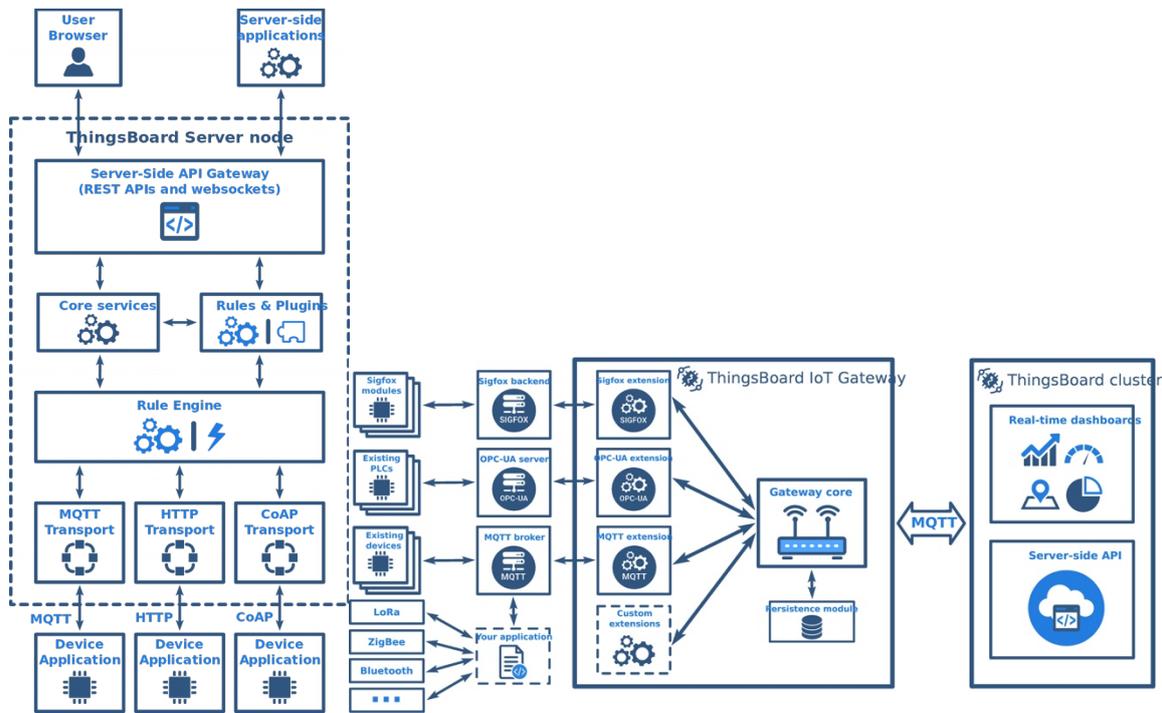

*Figure 3: ThingsBoard IoT Platform: architecture (left); and an example of the ThingsBoard IoT Gateway for Sigfox devices (right). (ThingsBoard, Inc. https://thingsboard.io.)*

**ThingsBoard Community Edition** [20] is an open-source IoT platform available from a GitHub repository (https://github.com/thingsboard/thingsboard) under the Apache License version 2.0. The company behind the platform also offers a commercial ``professional edition'' with additional support and extra platform integrations. The general architecture of ThingsBoard is shown in Fig. 3 (left). Connectivity with devices is handled via different transport components. In addition to the IoT platform there is also the ThingsBoard IoT Gateway to integrate IoT devices connected to third-party systems with ThingsBoard. An example of the usage of the IoT Gateway can be seen in Fig. 3 (right). Messages received are handled by the rule engine, which allows for both the processing of the data and triggering external alerts based on the content of the message. Data can be stored in an external PostgreSQL or Cassandra database. ThingsBoard utilizes Apache Zookeeper for cluster coordination and Cassandra as a NoSQL database. The core services are responsible for the device management, user management and dashboards. The server-side API Gateway provides a REST gateway that allows access to time-series data, and also allows registered users to send commands to devices. ThingsBoard has a plug-in architecture that allows coupling to external components. Existing plug-ins for Apache Kafka and sending emails are available. Internally, ThingsBoard uses Akka for event-driven message processing.

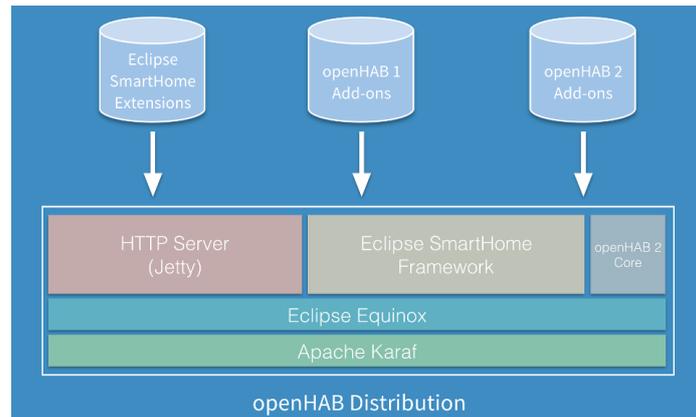
*Figure 4: openHAB 2 conceptual architecture. (openHAB Community, https://github.com/openhab.)*

**openHAB 2** [15] openHAB 2 is an open-source home automation platform, which is used for controlling and monitoring devices in the smart homes. It is licensed under Eclipse Public License 1.0, and uses a couple of Eclipse IoT projects (https://iot.eclipse.org/) mainly Eclipse SmartHome framework; see the reference architecture in Fig. 4. This platform has a well-documented, actively maintained GitHub repository (https://github.com/openhab), and provides an excellent support for variety of the smart devices. We had initial concerns about the applicability of a mobility use case, namely if there is a restriction on the number of smart devices that can be connected to the platform. It turns out scalability is not an issue; however, *the security component is entirely missing from the architecture design*, and it needs to be implemented from scratch. This is because of the intrinsic assumption that openHAB is used behind the home router firewall within one internal network. This ruled out the use of openHAB for our project.

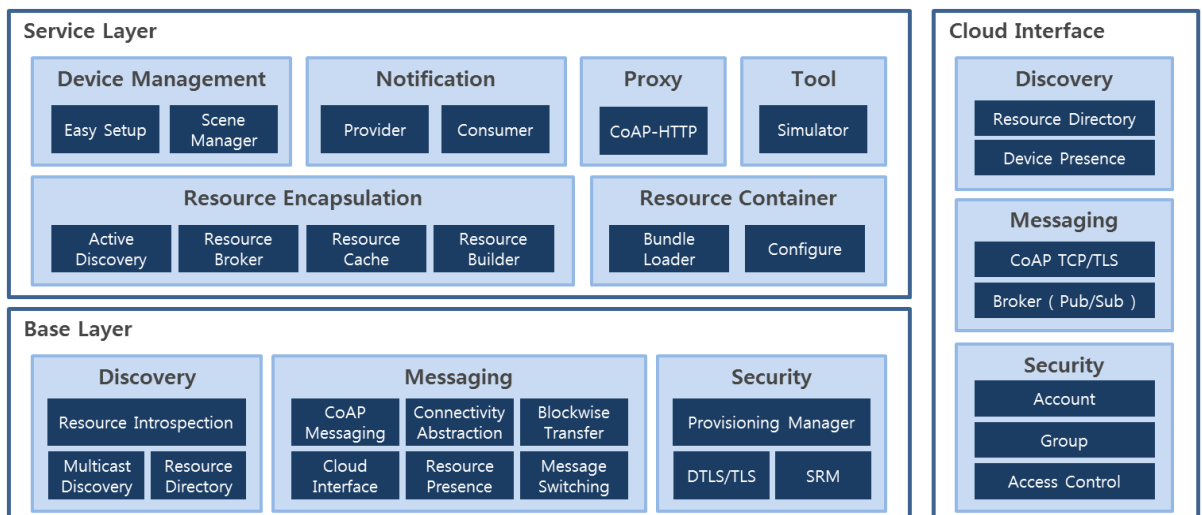
*Figure 5: IoTivity 1.2 Conceptual Architecture. (Open Interconnect Consortium, https://iotivity.org.)*

**IoTivity** [3] started out as a device management platform which enables seamless connectivity between devices. Upon merging it with another project AllJoyn, Open Connectivity Foundation (OCF) defined the purpose of IoTivity to be a set of ``specifications by OCF to ensure interoperability among connected devices'', as well as ``a reference implementation of the OCF specifications to the open-source community''.

IoTivity is an active project with the source code available via the GitHub repository ([https://github.com/iotivity](https://github.com/iotivity)). Similar to FiWARE, IoTivity offers cloud interface at the external gateway, it also supports discovery, messaging and security services within its base layer. This can be used to integrate the platform components with third-party systems. (For more details on the cloud part, we refer to one of the recent publications [4].) It is worth to mention that the platform provides a tool called *Simulator* which can help developers test their implementations without purchasing real hardware. The project also offers software components for the IoT device side for handshaking, resource registration/discovery, etc. The conceptual architecture of IoTivity is depicted in Fig. 5, and more details on the functionality of each component can be found at [https://wiki.iotivity.org/architecture](https://wiki.iotivity.org/architecture). There are Docker containers to ease the setup procedure, and it can be installed on various Linux distributions and Android system.

## 4 Conclusions

This project wanted to investigate the possible IoT platforms and look at the different features offered by each platform. In general, the choice of a suitable platform depends on the applications (use cases) researchers are trying to serve. We identified Kaa and ThingsBoard as candidate solutions based on the following criteria: permissible license, an actively maintained GitHub repository, clear architecture, and good documentation. However, if the aim is to have multiple applications served by the IoT platform, then it is best to start with a generic framework for interoperability reasons. In this case, the best suited platforms are IoTivity and FiWARE (e.g., smart city use case [13]), although they might require more effort in the implementation.

There are a lot of active developments in this field that researchers should to be aware of. For instance, Eclipse has a few IoT projects ([https://iot.eclipse.org/](https://iot.eclipse.org/)), which look promising including Eclipse Agail [6], and other open-source projects that have been reviewed by the recently published technical report [21].

There are advantages and disadvantages for using cloud-centric or self-maintained solutions. The self-maintained platform requires the presence of dedicated servers and an administrator maintaining the setup, connection, and is responsible for the backup. Finding a good hosting platform for the self-maintained systems is also a challenge. When arranging this on-premise, inside the academic institution, this will require collaboration with the centralized ICT services in the institute. Since the platform has strong requirements for the external network availability and access, centralized ICT might be hesitant in supporting it. External hosting providers will bring additional costs and risks concerning data security and availability.

In contrast, a cloud-centric solution comes with the cost determined by the service provider but will fully bypass the aforementioned issues. However, this type of solution means a full dependency on the service provider, including any changes to the API and service costs.

## Acknowledgements
*This work was jointly supported by SURF and the Netherlands eScience Center as the project City Cloud – From the Things to the Cloud and back, under the Enlighten Your Research-alliance program.*

Recent Advances, Taxonomy, Requirements, and Open Challenges. *IEEE Wireless Communications*, pages 10–16, 2017.